\documentclass[article,nojss]{jss}
\usepackage{amsbsy} 
\usepackage{amsmath}
\usepackage{amsfonts}
\usepackage{thumbpdf}
\usepackage{subfigure}
\usepackage{tabularx}
\usepackage{float}
\usepackage{graphicx}
\usepackage{ctable}
\usepackage{epstopdf}
\usepackage[latin1]{inputenc}
\newcommand{\real}{I\hspace{-1.1mm}R}

\def\bx{\boldsymbol{x}}

\def\bv{\boldsymbol{v}}
\def\bz{\boldsymbol{z}}

\def\bmu{\boldsymbol{\mu}}

\def\bSigma{\boldsymbol{\Sigma}}

\def\bpsi{\boldsymbol{\psi}}

\author{
	Antonio Punzo \\ University of Catania  
	\And
	Angelo Mazza \\ University of Catania
	\And
	Paul D. McNicholas \\ McMaster University
}    
\title{\pkg{ContaminatedMixt}: An \proglang{R} Package for Fitting Parsimonious Mixtures of Multivariate Contaminated Normal Distributions}

\Plainauthor{Antonio Punzo, Angelo Mazza, Paul D. McNicholas} 
\Plaintitle{ContaminatedMixt: An R Package for Fitting Parsimonious Mixtures of Multivariate Contaminated Normal Distributions} 
\Shorttitle{\pkg{ContaminatedMixt}: Parsimonious Mixtures of Contaminated Normal Distributions} 

\Abstract{
We introduce the \proglang{R} package \pkg{ContaminatedMixt}, conceived to disseminate the use of mixtures of multivariate contaminated normal distributions as a tool for robust clustering and classification under the common assumption of elliptically contoured groups.
Thirteen variants of the model are also implemented to introduce parsimony.
The expectation-conditional maximization algorithm is adopted to obtain maximum likelihood parameter estimates, and likelihood-based model selection criteria are used to select the model and the number of groups.
Parallel computation can be used on multicore PCs and computer clusters, when several models have to be fitted.
Differently from the more popular mixtures of multivariate normal and $t$ distributions, this approach also allows for automatic detection of mild outliers via the maximum \textit{a~posteriori} probabilities procedure.   
To exemplify the use of the package, applications to artificial and real data are presented.
}
\Keywords{mixture models, EM algorithm, contaminated normal distribution, outlier detection, robust clustering, robust estimates}

\Address{
  Antonio Punzo\\
  Department of Economics and Business\\
  University of Catania\\
  Corso Italia, 55, 95129 Catania, Italy\\
  Telephone: +39/095/7537640\\
  E-mail: \email{antonio.punzo@unict.it}\\
  URL: \url{http://www.economia.unict.it/punzo}\\ 

  Angelo Mazza\\
  Department of Economics and Business\\
  University of Catania\\
  Corso Italia, 55, 95129 Catania, Italy\\
  Telephone: +39/095/7537736\\
  E-mail: \email{a.mazza@unict.it}\\
  URL: \url{http://docenti.unict.it/a.mazza}\\ 
  
  Paul D. McNicholas\\
  Department of Mathematics \& Statistics\\
  McMaster University\\
  Hamilton, Ontario, Canada, L8S 4L8\\
  Telephone: +1-905-525-9140, ext.\ 23419\\
  E-mail: \email{mcnicholas@math.mcmaster.ca}\\
  URL: \url{http://ms.mcmaster.ca/~paul/} 
	
}

\begin{document}


\section{Introduction}
\label{sec:Introduction}

Finite mixtures of distributions are commonly used in statistical modelling 
as a powerful device for clustering and classification by often assuming that each mixture component represents a cluster (or group or class) into the original data (see \citealp{McLa:Basf:mixt:1988}, \citealp{Fral:Raft:Howm:1998}, and \citealp{Bohn:Comp:2000}).

For continuous multivariate random variables, attention is commonly focused on mixtures of multivariate normal distributions
because of their computational and theoretical convenience.
However, real data are often ``contaminated'' by outliers (also referred to as bad points herein, in analogy with \citealp{Aitk:Wils:Mixt:1980}) that affect the estimation of the component means and covariance matrices (see, e.g., \citealp{Barn:Lewi:Outl:1994}, \citealp{Beck:Gath:Them:1999}, \citealp{Bock:Clus:2002}, and \citealp{Gall:Ritt:Trim:2009}).

When outliers are mild \citep[see][for details]{Ritt:Robu:2015}, they can be dealt with by using heavy-tailed, usually elliptically symmetric, multivariate distributions.
Endowed with heavy tails, these distributions offer the flexibility needed for achieving mild outliers robustness, whereas the multivariate normal distribution, often used as the reference distribution for the good observations, lacks sufficient fit; for a discussion about the concept of reference distribution, see \citet{Davi:Gath:Thei:1993} and \citet{Henn:Fixe:2002}.
In this context, the multivariate $t$ distribution \citep[see, e.g.,][]{Lang:Litt:Tayl:Robu:1989} and the heavy-tailed versions of the multivariate power exponential distribution \citep[see, e.g.,][]{Gome:Gome:Main:Nava:Thee:2011} play a special role.
When used as mixture components, these distributions respectively yield mixtures of multivariate $t$ distributions (\citealp{McLa:Peel:Robu:1998} and \citealp{Peel:McLa:Robu:2000}) and mixtures of multivariate power exponential distributions \citep{Zhan:Lian:Robu:2010}.
Although these methods robustify the estimation of the component means and covariance matrices with respect to mixtures of multivariate normal distributions, they do not allow for automatic detection of bad points.
To overcome this problem, \citet{Punz:McNi:Robu:2015} introduce mixtures of multivariate contaminated normal distributions.
The multivariate contaminated normal distribution, which dates back to the seminal work of \citet{Tuke:Asur:1960}, is a further common and simple elliptically symmetric generalization of the multivariate normal distribution having heavier tails for the occurrence of bad points; it is a two-component normal mixture in which one of the components, with a large prior probability, represents the good observations (reference distribution), and the other, with a small prior probability, the same mean, and an inflated covariance matrix, represents the bad observations \citep{Aitk:Wils:Mixt:1980}. 
For further recent uses of this distribution in model-based clustering, see \citet{Punz:McNi:RCWM:2014,Punz:McNi:Robu:2014}, \citet{Punz:Maru:Clus:2016}, and \citet{Maru:Punz:Mode:2016}. 

In this paper we present the \proglang{R} \citep{R:2015} package \pkg{ContaminatedMixt}, available from CRAN at \url{https://cran.r-project.org/web/package=ContaminatedMixt}, which allows for model-based clustering and classification by means of a family, proposed by \citet{Punz:McNi:Robu:2015}, of fourteen parsimonious variants of mixtures of multivariate contaminated normal distributions.
Parsimony is attained by applying the eigen decomposition of the component scale matrices, in the fashion of \citet{Banf:Raft:mode:1993}.  
Fitting is performed via the expectation-conditional maximization (ECM) algorithm \citep{Meng:Rubin:Maxi:1993} and likelihood-based model selection criteria are adopted to select both the number of mixture components and the parsimonious model.

Several CRAN packages are available supporting model-based clustering and classification via mixtures of elliptically contoured distributions. 
A list of them may be found in the task view ``Cluster Analysis \& Finite Mixture Models'' of \citet{Leis:Grun:CRAN:2015}. 
One of the most flexible packages for clustering via mixtures of multivariate normal distributions is \pkg{mclust} (\citealp{Fral:Raft:Mode:2007} and \citealp{Fral:Raft:Scru:mclu:2015}); it provides ten of the fourteen parsimonious mixtures of multivariate normal distributions of \citet{Cele:Gova:Gaus:1995}, obtained via a slightly different eigen-decomposition of the component covariance matrices with respect to \citet{Banf:Raft:mode:1993}, implements an EM algorithm for model fitting, and uses the Bayesian information criterion \citep[BIC,][]{Schw:Esti:1978} to determine the number of components.
The package \pkg{Rmixmod} \citep{Lebr:Iovl:Lang:Bier:Cele:Gova:Rmix:2015} further fits the remaining four parsimonious models of \citet{Cele:Gova:Gaus:1995}.
The package \pkg{mixture} \citep{Brow:McNi:mixt:2015} allows to fit the family of fourteen parsimonious models of \citet{Banf:Raft:mode:1993}. 
Mixtures of multivariate normal distributions, with alternative parsimonious covariance structures, are also implemented by the packages \pkg{bgmm} \citep{Biec:Szcz:Ving:Tiur:TheR:2012} and \pkg{pgmm} \citep{McNi:Jamp:McDa:Murp:Bank:pgmm:2011}.   
The \pkg{teigen} package \citep{Andr:McNi:teig:2015} allows to fit a family of fourteen parsimonious mixtures of multivariate $t$-distributions (with eigen-decomposed component scale matrices as in \citealp{Cele:Gova:Gaus:1995}) from a clustering or classification point of view (see \citealp{Andr:McNi:Sube:Mode:2011} and \citealp{Andr:McNi:Mode:2012} for details).
Finally, although not available on CRAN, the \pkg{MPE} package, available at \url{http://onlinelibrary.wiley.com/doi/10.1111/biom.12351/suppinfo}, allows to fit a family, introduced by \citet{Dang:Brow:McNi:Mixt:2015}, of eight parsimonious variants of mixtures of multivariate power exponential distributions (with eigen-decomposed component scale matrices as in \citealp{Cele:Gova:Gaus:1995}).

The paper is organized as follows.
Section~\ref{sec:Methodology} retraces the models implemented in the \pkg{ContaminatedMixt} package, Section~\ref{sec:Maximum likelihood estimation} outlines the ECM algorithm for maximum likelihood parameters estimation, and Section~\ref{sec:fas} illustrates some further computational/practical aspects.
The relevance of the package is shown, via real and artificial data sets, in Section~\ref{sec:Package description and illustrative examples}, and conclusions are finally given in Section~\ref{sec:conclusions}.

\section{Methodology}
\label{sec:Methodology}

\subsection{The general model}
\label{subsec:The general model}

For a random vector $\boldsymbol{X}$, taking values in $\real^p$, a finite mixture of multivariate contaminated normal distributions \citep{Punz:McNi:Robu:2015} can be written as
\begin{equation}
p\left(\bx;\bpsi\right)=\sum_{g=1}^G\pi_g\left[\alpha_g\phi\left(\bx;\bmu_g,\boldsymbol{\Sigma}_g\right)+\left(1-\alpha_g\right)\phi\left(\boldsymbol{x};\boldsymbol{\mu}_g,\eta_g\boldsymbol{\Sigma}_g\right)\right],
\label{eq:mixture of multivariate contaminated normal densities}
\end{equation}
where, for the $g$th component, $\pi_g$ is its mixing proportion, with $\pi_g>0$ and $\sum_{g=1}^G\pi_g=1$, $\alpha_g\in\left(0,1\right)$ is the proportion of good observations, and $\eta_g>1$ denotes the degree of contamination.
In \eqref{eq:mixture of multivariate contaminated normal densities}, $\bpsi$ contains all of the parameters of the mixture while $\phi\left(\bx;\bmu,\bSigma\right)$ represents the distribution of a $p$-variate normal random vector with mean $\boldsymbol{\mu}$ and covariance matrix $\boldsymbol{\Sigma}$. 
As a special case, when $\alpha_g \rightarrow 1^-$ and $\eta_g \rightarrow 1^+$, for each $g=1,\ldots,G$, we obtain classical mixtures of multivariate normal distributions.

\subsection{Parsimonious variants of the general model}
\label{subsec:Parsimonious variants of the general model}

Because there are $p\left(p + 1\right)/2$ free parameters for each component scale matrix $\boldsymbol{\Sigma}_g$, it is usually necessary to introduce parsimony in model \eqref{eq:mixture of multivariate contaminated normal densities}. 
Following \citet{Banf:Raft:mode:1993}, \citet{Punz:McNi:Robu:2015} consider the eigen decomposition 
\begin{equation}
\bSigma_g=\lambda_g\boldsymbol{\Gamma}_g\boldsymbol{\Delta}_g\boldsymbol{\Gamma}_g^\top,
\label{eq:eigenvalue decomposition}
\end{equation}
where $\lambda_g$ is the first (largest) eigenvalue of $\bSigma_g$, $\boldsymbol{\Delta}_g$ is the diagonal matrix of the scaled (with respect to $\lambda_g$) eigenvalues of $\boldsymbol{\Sigma}_g$ sorted in decreasing order, and $\boldsymbol{\Gamma}_g$ is a $p\times p$ orthogonal matrix whose columns are the normalized eigenvectors of $\bSigma_g$, ordered according to their eigenvalues.
Each element in the right-hand side of \eqref{eq:eigenvalue decomposition} has a different geometric interpretation: $\lambda_g$ determines the size of the cluster, $\boldsymbol{\Delta}_g$ its shape, and $\boldsymbol{\Gamma}_g$ its orientation.

Following \citet{Banf:Raft:mode:1993}, \citet{Punz:McNi:Robu:2015} impose constraints on the three components of \eqref{eq:eigenvalue decomposition} resulting in a family of fourteen parsimonious mixtures of multivariate contaminated normal distributions (\tablename~\ref{tab:models}).
Sufficient conditions for the identifiability of the models in this family are given in \citet{Punz:McNi:Robu:2015}.
\begin{table*}[!ht]
\centering
\resizebox*{1\textwidth}{!}{
\begin{tabular}{lllllrr}
\toprule
Family & Model & Volume & Shape & Orientation & $\boldsymbol{\Sigma}_g$ &  
\# of free parameters in $\boldsymbol{\Sigma}_g$
\\
\midrule
Spherical & EII & Equal    & Spherical & --            & $\lambda \boldsymbol{I}$  & 1\\[1mm]
&VII & Variable & Spherical & --            & $\lambda_g \boldsymbol{I}$ & $G$\\[2mm]
Diagonal & EEI & Equal    & Equal     & Axis-Align & $\lambda \boldsymbol{\Gamma}$ & $p$\\[1mm]
&VEI & Variable & Equal     & Axis-Align & $\lambda_g \boldsymbol{\Gamma}$ & $G+p-1$\\[1mm]
&EVI & Equal    & Variable  & Axis-Align & $\lambda \boldsymbol{\Gamma}_g$ & $1+G\left(p-1\right)$\\[1mm]
&VVI & Variable & Variable  & Axis-Align & $\lambda_g \boldsymbol{\Gamma}_g$ & $Gp$\\[2mm]
General &EEE & Equal    & Equal     & Equal        & $\lambda\boldsymbol{\Gamma}\boldsymbol{\Delta}\boldsymbol{\Gamma}^\top$ & $p\left(p+1\right)/2$\\[1mm]
&VEE & Variable & Equal     & Equal        & $\lambda_g\boldsymbol{\Gamma}\boldsymbol{\Delta}\boldsymbol{\Gamma}^\top$ & $G+p-1+p\left(p-1\right)/2$ \\[1mm]
&EVE & Equal    & Variable  & Equal        & $\lambda\boldsymbol{\Gamma}\boldsymbol{\Delta}_g\boldsymbol{\Gamma}^\top$  & $1+G\left(p-1\right)+p\left(p-1\right)/2$ \\[1mm]
&EEV & Equal    & Equal     & Variable     & $\lambda\boldsymbol{\Gamma}_g\boldsymbol{\Delta}\boldsymbol{\Gamma}_g^\top$ &  $p+Gp\left(p-1\right)/2$ \\[1mm]
&VVE & Variable & Variable  & Equal        & $\lambda_g\boldsymbol{\Gamma}\boldsymbol{\Delta}_g\boldsymbol{\Gamma}^\top$ & $Gp+p\left(p-1\right)/2$ \\[1mm]
&VEV & Variable & Equal     & Variable   & $\lambda_g\boldsymbol{\Gamma}_g\boldsymbol{\Delta}\boldsymbol{\Gamma}_g^\top$ & $G+p-1+Gp\left(p-1\right)/2$ \\[1mm]
&EVV & Equal    & Variable  & Variable   & $\lambda\boldsymbol{\Gamma}_g\boldsymbol{\Delta}_g\boldsymbol{\Gamma}_g^\top$ & $1+G\left(p-1\right)+Gp\left(p-1\right)/2$ \\[1mm]
&VVV & Variable & Variable  & Variable   & $\lambda_g\boldsymbol{\Gamma}_g\boldsymbol{\Delta}_g\boldsymbol{\Gamma}_g^\top$ & $Gp\left(p+1\right)/2$ \\
\bottomrule
\end{tabular}
}
\caption{
\label{tab:models}
Nomenclature, covariance structure, and number of free parameters in $\boldsymbol{\Sigma}_g$ for each member of the family of parsimonious mixtures of multivariate contaminated normal distributions.}
\end{table*}

\subsection{Modelling framework: model-based classification}
\label{subsec:modelling frame: model-based classification}

Model-based classification is receiving renewed attention (see, e.g., \citealp{Dean:Murp:Down:2006}, \citealp{McNi:Mode:2010}, \citealp{Andr:McNi:Sube:Mode:2011}, \citealp{Brow:McNi:Mode:2012}, \citealp{Punz:Flex:2014}, and \citealp{Sube:Punz:Ingr:McNi:Clus:2013,Sube:Punz:Ingr:McNi:Clus:2015}).
However, despite being the most general framework within which to present and analyze direct applications of mixture models, it remains the ``poor cousin'' of model-based clustering within the literature.

Consider the random sample $\left\{\boldsymbol{x}_i\right\}_{i=1}^n$ from \eqref{eq:mixture of multivariate contaminated normal densities}.
Without loss of generality, suppose that the first $m$ observations are known to belong to one of $G$ groups; these are the so-called labeled observations. 
Let $\boldsymbol{z}_i$ be the $G$-dimensional component-label vector in which the $g$th element is $z_{ig}=1$ if $\boldsymbol{x}_i$ belongs to component $g$ and $z_{ig}=0$ otherwise, $g=1,\ldots,G$.
If the $i$th observation is labeled, denote with $\widetilde{\bz}_i=\left(\widetilde{z}_{i1},\ldots,\widetilde{z}_{iG}\right)$ its component-membership indicator. 
In model-based classification, we use all $n$ observations to estimate the parameters of the mixture; the fitted model is adopted to classify each of the $n-m$ unlabeled observations through the corresponding maximum \textit{a posteriori} (MAP) probability.
Note that 
$$
\text{MAP}\left(z_{ig}\right)=
\begin{cases}
1 & \text{if } \max_h\{z_{ih}\} \text{ occurs in component }g,\\
0 & \text{otherwise}.
\end{cases}
$$
Using this notation, the model-based classification likelihood can be written as
\begin{displaymath}
\mathcal{L}\left(\bpsi\right)=\prod_{i=1}^m\prod_{g=1}^G\left\{\pi_g\left[\alpha_g\phi\left(\boldsymbol{x}_i;\boldsymbol{\mu}_g,\boldsymbol{\Sigma}_g\right)+\left(1-\alpha_g\right)\phi\left(\boldsymbol{x}_i;\boldsymbol{\mu}_g,\eta_g\boldsymbol{\Sigma}_g\right)\right]\right\}^{\widetilde{z}_{ig}} \prod_{i=m+1}^n p\left(\boldsymbol{x}_i;\bpsi\right).
\end{displaymath}
We obtain the model-based clustering scenario as a special case when $m=0$; this is the scenario used by \citet{Punz:McNi:Robu:2015} to introduce the model.

\section{Maximum likelihood estimation}
\label{sec:Maximum likelihood estimation}

\subsection{An ECM algorithm}
\label{subsec:An ECM Algorithm}

To fit the models in \tablename~\ref{tab:models}, \citet{Punz:McNi:Robu:2015} illustrate the expectation-conditional maximization (ECM) algorithm of \citet{Meng:Rubin:Maxi:1993}. 
The ECM algorithm is a variant of the classical expectation-maximization (EM) algorithm \citep{Demp:Lair:Rubi:Maxi:1977}, which is a natural approach for maximum likelihood estimation when data are incomplete. 
In our case, there are two sources of missing data: one arises from the fact that we do not know the component labels $\left\{\boldsymbol{z}_i\right\}_{i=m+1}^n$ and the other arises from the fact that we do not know whether an observation in group $g$ is good or bad. 
To denote this second source of missing data, we use $\left\{\boldsymbol{v}_i\right\}_{i=1}^n$, with $\boldsymbol{v}_i=(v_{i1},\ldots,v_{iG})$, where $v_{ig}=1$ if observation~$i$ in group~$g$ is good and $v_{ig}=0$ if observation $i$ in group $g$ is bad.
By working on the complete-data likelihood 
\begin{eqnarray}
\mathcal{L}_c\left(\bpsi\right)&=&\prod_{i=1}^m\prod_{g=1}^G\left\{\pi_g\left[\alpha_g\phi\left(\boldsymbol{x}_i;\boldsymbol{\mu}_g,\boldsymbol{\Sigma}_g\right)\right]^{v_{ig}}\left[\left(1-\alpha_g\right)\phi\left(\boldsymbol{x}_i;\boldsymbol{\mu}_g,\eta_g\boldsymbol{\Sigma}_g\right)\right]^{\left(1-v_{ig}\right)}\right\}^{\widetilde{z}_{ig}} \nonumber\\
&& \times \prod_{i=m+1}^n\prod_{g=1}^G\left\{\pi_g\left[\alpha_g\phi\left(\boldsymbol{x}_i;\boldsymbol{\mu}_g,\boldsymbol{\Sigma}_g\right)\right]^{v_{ig}}\left[\left(1-\alpha_g\right)\phi\left(\boldsymbol{x}_i;\boldsymbol{\mu}_g,\eta_g\boldsymbol{\Sigma}_g\right)\right]^{\left(1-v_{ig}\right)}\right\}^{z_{ig}},
\label{eq:complete-data likelihood}
\end{eqnarray}
the ECM algorithm iterates between three steps --- an E-step and two CM-steps --- until convergence (which is evaluated via the Aitken acceleration criterion; see \citealp{Aitk:OnBe:1926} and \citealp{Lind:Mixt:1995}).
The only difference from the EM algorithm is that each M-step is replaced by two simpler CM-steps. 
They arise from the partition $\bpsi=\left\{\boldsymbol{\psi}_1,\boldsymbol{\psi}_2\right\}$, where $\boldsymbol{\psi}_1=\left\{\pi_g,\alpha_g,\boldsymbol{\mu}_g,\boldsymbol{\Sigma}_g\right\}_{g=1}^G$ and $\boldsymbol{\psi}_2=\left\{\eta_g\right\}_{g=1}^G$.
In particular, for the most general model VVV, the $\left(r+1\right)$th iteration of the ECM algorithm can be summarized/simplified as follows \citep[see][for details on the model-based clustering paradigm]{Punz:McNi:Robu:2015}:
\begin{description}
	\item[E-step:] The values of $z_{ig}$ and $v_{ig}$ in \eqref{eq:complete-data likelihood} are respectively replaced by
\begin{equation*}
z_{ig}^{\left(r\right)}=\frac{\pi_g^{\left(r\right)}\left[\alpha_g^{\left(r\right)}\phi\left(\boldsymbol{x};\boldsymbol{\mu}_g^{\left(r\right)},\boldsymbol{\Sigma}_g^{\left(r\right)}\right)+\left(1-\alpha_g^{\left(r\right)}\right)\phi\left(\boldsymbol{x};\boldsymbol{\mu}_g^{\left(r\right)},\eta_g^{\left(r\right)}\boldsymbol{\Sigma}_g^{\left(r\right)}\right)\right]}{p\left(\boldsymbol{x};\bpsi^{\left(r\right)}\right)}
\end{equation*}
and 
\begin{equation*}
v_{ig}^{\left(r\right)}=\frac{\alpha_g^{\left(r\right)}\phi\left(\boldsymbol{x};\boldsymbol{\mu}_g^{\left(r\right)},\boldsymbol{\Sigma}_g^{\left(r\right)}\right)}{\alpha_g^{\left(r\right)}\phi\left(\boldsymbol{x};\boldsymbol{\mu}_g^{\left(r\right)},\boldsymbol{\Sigma}_g^{\left(r\right)}\right)+\left(1-\alpha_g^{\left(r\right)}\right)\phi\left(\boldsymbol{x};\boldsymbol{\mu}_g^{\left(r\right)},\eta_g^{\left(r\right)}\boldsymbol{\Sigma}_g^{\left(r\right)}\right)};
\end{equation*}
	\item[CM-step 1:]
	Fixed $\boldsymbol{\psi}_2=\boldsymbol{\psi}_2^{\left(r\right)}$, the parameters in $\boldsymbol{\psi}_1$ are updated as
\begin{displaymath}
\pi_g^{\left(r+1\right)}=\frac{n_g^{\left(r\right)}}{n},
\end{displaymath}
\begin{equation}
\alpha_g^{\left(r+1\right)}=\frac{1}{n_g^{\left(r\right)}}\left(\sum_{i=1}^m\widetilde{z}_{ig}v_{ig}^{\left(r\right)}+\sum_{i=m+1}^nz_{ig}^{\left(r\right)}v_{ig}^{\left(r\right)}\right),
\label{eq:alpha}
\end{equation}
\begin{equation}
\displaystyle\boldsymbol{\mu}_g^{\left(r+1\right)}=\frac{1}{s_g^{\left(r\right)}}\Biggl[\sum_{i=1}^m\widetilde{z}_{ig}\left(v_{ig}^{\left(r\right)}+\frac{1-v_{ig}^{\left(r\right)}}{\eta_g^{\left(r\right)}}\right)\boldsymbol{x}_i+\sum_{i=m+1}^nz_{ig}^{\left(r\right)}\left(v_{ig}^{\left(r\right)}+\frac{1-v_{ig}^{\left(r\right)}}{\eta_g^{\left(r\right)}}\right)\boldsymbol{x}_i\Biggr],
\label{eq:mu decomposition}
\end{equation}
and
\begin{equation*}
\displaystyle\boldsymbol{\Sigma}_g^{\left(r+1\right)}=\frac{1}{n_g^{\left(r\right)}}\boldsymbol{W}_g^{\left(r\right)},
\end{equation*}
where
\begin{equation*}
n_g^{\left(r\right)}=\sum_{i=1}^m\widetilde{z}_{ig}+\sum_{i=m+1}^nz_{ig}^{\left(r\right)},
\end{equation*}
\begin{equation*}
s_g^{\left(r\right)}=\sum_{i=1}^m\widetilde{z}_{ig}\left(v_{ig}^{\left(r\right)}+\frac{1-v_{ig}^{\left(r\right)}}{\eta_g^{\left(r\right)}}\right)+\sum_{i=m+1}^nz_{ig}^{\left(r\right)}\left(v_{ig}^{\left(r\right)}+\frac{1-v_{ig}^{\left(r\right)}}{\eta_g^{\left(r\right)}}\right),
\end{equation*}
and
\begin{eqnarray}
\boldsymbol{W}_g^{\left(r+1\right)}&=&\sum_{i=1}^m\widetilde{z}_{ig}\left(v_{ig}^{\left(r\right)}+\frac{1-v_{ig}^{\left(r\right)}}{\eta_g^{\left(r\right)}}\right)\left(\boldsymbol{x}_i-\displaystyle\boldsymbol{\mu}_g^{\left(r+1\right)}\right)\left(\boldsymbol{x}_i-\displaystyle\boldsymbol{\mu}_g^{\left(r+1\right)}\right)^\top\nonumber\\
&&+\sum_{i=m+1}^nz_{ig}^{\left(r\right)}\left(v_{ig}^{\left(r\right)}+\frac{1-v_{ig}^{\left(r\right)}}{\eta_g^{\left(r\right)}}\right)\left(\boldsymbol{x}_i-\displaystyle\boldsymbol{\mu}_g^{\left(r+1\right)}\right)\left(\boldsymbol{x}_i-\displaystyle\boldsymbol{\mu}_g^{\left(r+1\right)}\right)^\top.
\label{eq:W decomposition}
\end{eqnarray}
	\item[CM-step 2:] Fixed $\boldsymbol{\psi}_1=\boldsymbol{\psi}_1^{\left(r+1\right)}$, the parameters in $\boldsymbol{\psi}_2$ are updated by maximizing the function
	\begin{equation}
\begin{array}{l}
\displaystyle-\frac{p}{2}\sum_{i=1}^m\widetilde{z}_{ig}\left(1-v_{ig}^{\left(r\right)}\right)\ln \eta_g
\displaystyle-\frac{p}{2}\sum_{i=m+1}^nz_{ig}^{\left(r\right)}\left(1-v_{ig}^{\left(r\right)}\right)\ln \eta_g + \\[5mm]
\displaystyle-\frac{1}{2}\sum_{i=1}^m\widetilde{z}_{ig}\frac{1-v_{ig}^{\left(r\right)}}{\eta_g}
\delta\left(\boldsymbol{x}_i,\boldsymbol{\mu}_g^{\left(r+1\right)};\boldsymbol{\Sigma}_g^{\left(r+1\right)}\right)-\frac{1}{2}\sum_{i=m+1}^n\widetilde{z}_{ig}^{\left(r\right)}\frac{1-v_{ig}^{\left(r\right)}}{\eta_g}\delta\left(\boldsymbol{x}_i,\boldsymbol{\mu}_g^{\left(r+1\right)};\boldsymbol{\Sigma}_g^{\left(r+1\right)}\right),
\end{array}
\label{eq:maximization function for etaj}
\end{equation}
where $\delta\left(\bx_i,\bmu_g^{\left(r+1\right)};\bSigma_g^{\left(r+1\right)}\right)$ denotes the squared Mahalanobis distance between $\bx_i$ and $\bmu_g^{\left(r+1\right)}$ (with covariance matrix $\bSigma_g^{\left(r+1\right)}$), with respect to $\eta_g$, under the constraint $\eta_g>1$, for $g=1,\ldots,G$.
Operationally, the \code{optimize()} function, in the \pkg{stats} package, is used to perform a numerical search of the maximum $\eta_g^{\left(r+1\right)}$ of \eqref{eq:maximization function for etaj} over the interval $\left(1,\eta_g^*\right)$, with $\eta_g^*>1$.
\end{description}

As it is well-documented in \citet{Punz:McNi:Robu:2015}, the weights 
$$
\left(v_{ig}^{\left(r\right)}+\frac{1-v_{ig}^{\left(r\right)}}{\eta_g^{\left(r\right)}}\right)
$$
in \eqref{eq:mu decomposition} and \eqref{eq:W decomposition} reduce the impact of bad points in the estimation of the component means $\boldsymbol{\mu}_g$ and the component scale matrices $\boldsymbol{\Sigma}_g$, thereby providing robust estimates of these parameters.
For a discussion on down-weighting for the multivariate contaminated normal distribution, see also \citet{Litt:Robu:1988}. 

The ECM algorithm for the other parsimonious models changes only with respect to the way the terms of the decomposition of $\boldsymbol{\Sigma}_g$ are obtained in the first CM-step. 
In particular, these updates are analogous to those given by \citet{Cele:Gova:Gaus:1995} for their normal parsimonious clustering (GPC) models (corresponding to mixtures of multivariate normal distributions with eigen decomposed covariance matrices). 
The only difference is that, on the $\left(r+1\right)$th iteration of the algorithm, $\boldsymbol{W}_g^{\left(r+1\right)}$ is used instead of the classical scattering matrix 
\begin{equation*}
\sum_{i=1}^m\widetilde{z}_{ig}\left(\boldsymbol{x}_i-\displaystyle\boldsymbol{\mu}_g^{\left(r+1\right)}\right)\left(\boldsymbol{x}_i-\displaystyle\boldsymbol{\mu}_g^{\left(r+1\right)}\right)^\top+\sum_{i=m+1}^nz_{ig}^{\left(r\right)}\left(\boldsymbol{x}_i-\displaystyle\boldsymbol{\mu}_g^{\left(r+1\right)}\right)\left(\boldsymbol{x}_i-\displaystyle\boldsymbol{\mu}_g^{\left(r+1\right)}\right)^\top.	
\end{equation*}


\section{Further aspects}
\label{sec:fas}

\subsection[Initialization]{Initialization}
\label{subsec:Initialization}

Many initialization strategies have been proposed for the EM algorithm applied to mixture models (see, e.g., \citealp{Bier:Cele:Gova:Choo:2003}, \citealp{Karl:Xeka:Choo:2003}, and \citealp{Bagn:Punz:Fine:2013}).
The \pkg{ContaminatedMixt} package implements the following initializations, all based on providing the initial quantities $\boldsymbol{z}_{i}^{\left(0\right)}$, $\boldsymbol{v}_{i}^{\left(0\right)}$, and $\eta_g^{\left(0\right)}=1.001$, $i=1,\ldots,n$ and $g=1,\ldots,G$, to the first CM-step of the ECM algorithm.
\begin{description}
	\item[\code{"random.soft"}:] each $\boldsymbol{z}_{i}^{\left(0\right)}$ is substituted by a single observation randomly generated --- via the \code{rmultinom()} function of the \pkg{stats} package --- from a multinomial distribution with probabilities $\left(1/G,\ldots,1/G\right)$.	
The values $v_{ig}^{\left(0\right)}$, $i=1,\ldots,n$ and $g=1,\ldots,G$, are, by default, fixed to one, but they can be also provided by the user.   
	\item[\code{"random.hard"}:] the $G$ values in $\boldsymbol{z}_{i}^{\left(0\right)}$ are randomly generated by a uniform distribution --- via the \code{runif()} function of the \pkg{stats} package --- and then normalized in order to sum to 1.
	The values $v_{ig}^{\left(0\right)}$, $i=1,\ldots,n$ and $g=1,\ldots,G$, are, by default, fixed to one, but they can be also provided by the user.   
	\item[\code{"kmeans"}:] hard values for $\boldsymbol{z}_{i}^{\left(0\right)}$, $i=1,\ldots,n$, are provided by a preliminary run of the $k$-means algorithm, as implemented by the \code{kmeans()} function of the \pkg{stats} package.
	\item[\code{"mixt"}:] For each parsimonious model, the $n$ values $\boldsymbol{z}_{i}^{\left(0\right)}$ are substituted with the posterior probabilities arising from the fitting of the corresponding parsimonious mixture of multivariate normal distributions; the latter is estimated by the \code{gpcm()} function of the \pkg{mixture} package.
	The values $v_{ig}^{\left(0\right)}$, $i=1,\ldots,n$ and $g=1,\ldots,G$, are fixed to one;
	from an operational point of view, thanks to the monotonicity property of the ECM algorithm \citep[see, e.g.,][p.~33]{McLa:Kris:TheE:2007}, this also guarantees that the final observed-data log-likelihood of the parsimonious mixture of multivariate contaminated normal distributions will be always greater than, or equal to, the observed-data log-likelihood of the corresponding parsimonious mixture of multivariate normal distributions.
This is a fundamental consideration for the use of likelihood-based model selection criteria for choosing between these two models.   
	\item[\code{"manual"}:] the (soft or hard) values of $\boldsymbol{z}_{i}^{\left(0\right)}$, as well as the values of $\boldsymbol{v}_{i}^{\left(0\right)}$, are provided by the user.
\end{description}

%

\subsection{Automatic detection of bad points}
\label{subsec:Automatic detection of bad points}

For a mixture of multivariate contaminated normal distributions, the classification of an observation $\bx_i$ means: 
\begin{description}
	\item[Step 1.] to determine its cluster of membership;
	\item[Step 2.] to establish if it is either a good or a bad observation in that cluster.
\end{description}
Let $\widehat{\bz}_i$ and $\widehat{\bv}_i$ denote, respectively, the expected values of $\bz_i$ and $\bv_i$ arising from the ECM algorithm, i.e., $\widehat{z}_{ig}$ is the value of $z_{ig}^{\left(r\right)}$ at convergence and $\widehat{v}_{ig}$ is the value of $v_{ig}^{\left(r\right)}$ at convergence. 
To determine the cluster of membership of $\bx_i$, we use the MAP classification, i.e., $\text{MAP}\left(\widehat{z}_{ig}\right)$.
We then consider $\widehat{v}_{ih}$, where $h$ is selected such that $\text{MAP}\left(\widehat{z}_{ih}\right)=1$, while $\bx_i$ is considered good if $\widehat{v}_{ih}>0.5$ and $\bx_i$ is considered bad otherwise. 
The resulting information can be used to eliminate the bad points, if such an outcome is desired \citep{Berk:Bent:Esti:1988}. 
The remaining data may then be treated as effectively being distributed according to a mixture of multivariate normal distributions, and the clustering results can be reported as usual.

\subsection{Constraints for detection of bad points}
\label{subsec:Constraints for detection of outliers}

It may be required that in the $g$th cluster, $g=1,\ldots,G$, the proportion of good data is at least equal to a pre-determined value $\alpha_g^*$. 
In this case, the \texttt{optimize()} function is also used for a numerical search of the maximum $\alpha_g^{\left(r+1\right)}$, over the interval $\left(\alpha_g^*,1\right)$, of the function
\begin{equation}
\sum_{i=1}^m\widetilde{z}_{ig}^{\left(r\right)}\left[v_{ig}^{\left(r\right)}\ln \alpha_g+\left(1-v_{ig}^{\left(r\right)}\right)\ln \left(1-\alpha_g\right)\right]+\sum_{i=m+1}^nz_{ig}^{\left(r\right)}\left[v_{ig}^{\left(r\right)}\ln \alpha_g+\left(1-v_{ig}^{\left(r\right)}\right)\ln \left(1-\alpha_g\right)\right].	
\label{eq:second constraint}
\end{equation}
Note that the \pkg{ContaminatedMixt} package also allows to fix $\alpha_g$ \textit{a~priori}. 
This is somewhat analogous to the trimmed clustering approach implemented by the \pkg{tclust} package \citep{Frit:Garc:Mayo:tclust:2012}, where one must specify the proportion of outliers (the so-called trimming proportion) in advance. 
However, pre-specifying the proportion of bad points \textit{a~priori} may not be realistic in many practical scenarios.

\subsection{Model selection criteria}
\label{subsec:BIC}

Thus far, the number of components $G$ and the covariance structure (cf.~\tablename~\ref{tab:models}) have been treated as \textit{a~priori} fixed.
However, in most practical applications, they are unknown, so it is common practice to select them by evaluating a convenient (likelihood-based) model selection criterion over a reasonable range of possible options (for the alternative use of likelihood-ratio tests to select either the parsimonious model or the number of components for a normal mixture, see \citealp{Punz:Brow:McNi:Hypo:2016}).
The \pkg{ContaminatedMixt} package supports the information criteria listed in \tablename~\ref{tab:IC}, where $l\left(\widehat{\bpsi}\right)$ is the observed-data log-likelihood and $q$ is the number of free parameters.
\begin{table}[!ht]
\centering
\resizebox*{1\textwidth}{!}{
\begin{tabular}{*{3}c}
\toprule
information 
criterion
& 
definition 
&  
reference
\\
\midrule
AIC     & $2l\left(\widehat{\bpsi}\right)-2q$ & \citet{Akai:Info:1973} \\[3mm]
AIC$_3$ & $2l\left(\widehat{\bpsi}\right)-3q$ & \citet{Bozd:Theo:1974} \\[3mm]
AICc    & $\text{AIC}-2\displaystyle\frac{q\left(q+1\right)}{n-q-1}$ & \citet{Hurv:Tsai:Regr:1989}      \\[3mm]
AICu    & $\text{AICc}-n\ln\displaystyle\frac{n}{n-q-1}$                       & \citet{McQu:Shum:Tsai:Them:1997} \\[3mm]
AWE     & $2l\left(\widehat{\bpsi}\right)-2\rho\left(\displaystyle\frac{3}{2}+\ln n\right)$                & \citet{Banf:Raft:mode:1993}      \\[3mm]
BIC     & $2l\left(\widehat{\bpsi}\right)-q\ln n$                                                       & \citet{Schw:Esti:1978}           \\[3mm]
CAIC    & $2l\left(\widehat{\bpsi}\right)-q\left(1+\ln n\right)$                                        & \citet{Bozd:Mode:1987}           \\[3mm]
ICL     & $\text{BIC} + \displaystyle\sum_{i=m+1}^n\sum_{g=1}^G \text{MAP}\left(\widehat{z}_{ig}\right)\ln \widehat{z}_{ig}$ & \citet{Bier:Cele:Gova:Asse:2000} \\
\bottomrule
\end{tabular}
}
\caption{
\label{tab:IC}
Definition and key reference for the implemented model selection criteria.
}
\end{table}      

\section{Package description and illustrative example}
\label{sec:Package description and illustrative examples}

In this section we provide a description of the main capabilities of the \pkg{ContaminatedMixt} package along with some illustrations.

\subsection{Package description}
\label{subsec:Package description}

The \pkg{ContaminatedMixt} package is developed in an object-oriented design, using the standard S3 paradigm. 
Its main function, \code{CNmixt()}, fits the model(s) in \tablename~\ref{tab:models} and returns a \code{ContaminatedMixt} class object; the arguments of this function, along with their description, are listed in \tablename~\ref{tab:arguments}. 
\begin{table}[!ht]
\centering
\begin{tabularx}{\linewidth}{l  X}
\hline
arguments & description \\ 
\hline
		\code{X}     & matrix or data frame of dimension $n \times p$, with $p>1$. 
		\\[0.5mm]
		\code{G}     & vector containing the numbers of groups to be tried. 
		\\[0.5mm]
		\code{model} & vector indicating the models (\code{"EII"}, \code{"VII"}, \code{"EEI"}, \code{"VEI"}, \code{"EVI"}, \code{"VVI"}, \code{"EEE"}, \code{"VEE"}, \code{"EVE"}, \code{"EEV"}, \code{"VVE"}, \code{"VEV"}, \code{"EVV"}, \code{"VVV"}) to be used.
		If \code{model = NULL} (default), then all 14 models are fitted. 
		\\[0.5mm]
		\code{initialization}  & initialization strategy for the ECM algorithm. Possible values are \code{"random.soft"}, \code{"random.hard"}, \code{"kmeans"}, \code{"mixt"}, and \code{"manual"} (see Section~\ref{subsec:Initialization} for details).	
		Default is \code{initialization = "mixt"}. 
		\\[0.5mm]
		\code{alphafix} & vector, of dimension $G$, with fixed \textit{a~priori} values for $\alpha_1,\ldots,\alpha_G$.
		If the length of \code{alphafix} is different from $G$, its first element is replicated $G$ times.
		If \code{alphafix = NULL} (default), $\alpha_1,\ldots,\alpha_G$ are estimated.
		\\[0.5mm]
		\code{alphamin} & vector with values $\alpha_1^*,\ldots,\alpha_G^*$ (see Section~\ref{subsec:Constraints for detection of outliers}).
		If the length of \code{alphamin} is different from $G$, its first element is replicated $G$ times.
		If \code{alphamin = NULL}, $\alpha_1,\ldots,\alpha_G$ are estimated without constraints, as in \eqref{eq:alpha}.
		Default value is \code{0.5}.
		\\[0.5mm] 
		\code{etafix} & vector, of dimension $G$, with fixed \textit{a~priori} values for $\eta_1,\ldots,\eta_G$.
		If the length of \code{etafix} is different from $G$, its first element is replicated $G$ times.
		If \code{etafix = NULL} (default), $\eta_1,\ldots,\eta_G$ are estimated.
		\\[0.5mm]
		\code{etamax} & vector with values $\eta_1^*,\ldots,\eta_G^*$ (see the CM-step 2 of Section~\ref{subsec:An ECM Algorithm}).
		If the length of \code{etamax} is different from $G$, its first element is replicated $G$ times.
		Default value is \code{1000}.
		\\[0.5mm]
		\code{seed} & the seed for the random number generator, when random initializations are used; if \code{NULL} (default), current seed is not changed. 
		\\[0.5mm]
		\code{start.z} & when \code{initialization = "manual"}, it is a $n\times G$ matrix with values $z_{ig}^{(0)}$.
		\\[0.5mm]
		\code{start.v} & it is a $n\times G$ matrix with values $v_{ig}^{(0)}$.	
		If \code{start.v = NULL} (default), then $v_{ig}^{(0)}=1$, $i=1,\ldots,n$ and $g=1,\ldots,G$. 
		\\[0.5mm]
		\code{ind.label} & vector of positions (rows) of the labeled observations. 
		\\[0.5mm]
		\code{label}     & vector, of the same dimension of \code{ind.label}, with the groups membership of the observations indicated by \code{ind.label}. 
		\\[0.5mm]
		\code{iter.max} & maximum number of iterations in the ECM algorithm. 
		Default is \code{1000}. 
		\\[0.5mm]
		\code{threshold} & value of the stopping rule for Aitken's acceleration procedure.	
		Default is \code{1.0e-03}. 
		\\[0.5mm]
		\code{eps} & smallest value for the eigenvalues of $\bSigma_1,\ldots,\bSigma_G$.
		It used to prevent the EM algorithm to be affected by local maxima or degeneracy of the likelihood (see \citealp{Hath:Acons:1986} and \citealp{Ingr:Alik:2004}).
		Default value is \code{1e-100}. 
		\\[0.5mm]
		\code{parallel} & when \code{TRUE}, the package \pkg{parallel} is used for parallel computation.
		The number of cores to use may be set with the global option \code{cl.cores}; default value is detected using \code{detectCores()}. \\
\hline
\end{tabularx}
\caption{List of arguments for the function \code{CNmixt()}.}
\label{tab:arguments}
\end{table}
In addition, the package contains several methods that allow for data extraction, visualization, and plot.

Extractors for \code{ContaminatedMixt} class objects are illustrated in \tablename~\ref{tab:Extractors}.
\begin{table}[!ht]
\centering
\begin{tabularx}{\linewidth}{l  X}
\hline
extractors & description \\ 
\hline
	\code{getBestModel()}	& a \code{ContaminatedMixt} class object containing the best model only.\\[1.5mm]
	\code{getPosterior()}	& estimated posterior probabilities $\widehat{z}_{ig}$, $i=1,\ldots,n$ and $g=1,\ldots,G$.\\[1.5mm]
	\code{getSize()}	    & estimated groups sizes (from the hard classification induced by the MAP operator).\\[1.5mm]
	\code{getCluster()}	  & classification vector.\\[1.5mm]
	\code{getDetection()} & matrix with two columns: the first gives the MAP group memberships whereas the second specifies if the observations are either good or bad (see Section~\ref{subsec:Automatic detection of bad points}) \\[1.5mm]
	\code{getPar()}	      & estimated parameters (i.e., $\widehat{\bpsi}$).\\[1.5mm]
	\code{getIC()}	      & values for the considered criteria in \code{criteria}.\\[1.5mm]
	\code{whichBest()}	  & position of the model, in the \code{ContaminatedMixt} class object, for the criteria specified in \code{criteria}. \\
\hline
\end{tabularx}
\caption{Extractors for \code{ContaminatedMixt} class objects.}
\label{tab:Extractors}
\end{table}
When several models have been fitted, extractor functions consider the best model according to the information criterion in \code{criterion}, within the subset of estimated models having a number of components among those in \code{G} and a parsimonious model among those in \code{model}. 
Note that \code{getIC()} and \code{whichBest()} have an argument \code{criteria}, in substitution to \code{criterion}, which allows to select more than one criterion. 

The package also includes some methods for \code{ContaminatedMixt} class objects; they are: \code{plot()} and \code{pairs()}, to display clustering/classification results in terms of scatter plots (in the cases $p=2$ and $p\geq 2$, respectively), \code{summary()}, to visualize the estimated parameters and further inferential/clustering details, \code{print()}, to print at video the selected model(s) according to the information criteria in \tablename~\ref{tab:IC}, and \code{agree()} to evaluate the agreement of a given partition with respect to the partition arising from a fitted model. 
As usual, further details can be found in the functions' help pages.

Finally, for the multivariate contaminated normal distribution, the function \code{dCN()} gives the probability density and the function \code{rCN()} generates random deviates.


%

\subsection{Artificial data}
\label{subsec:Artificial data}

\begin{figure}[!ht]
\centering
\includegraphics{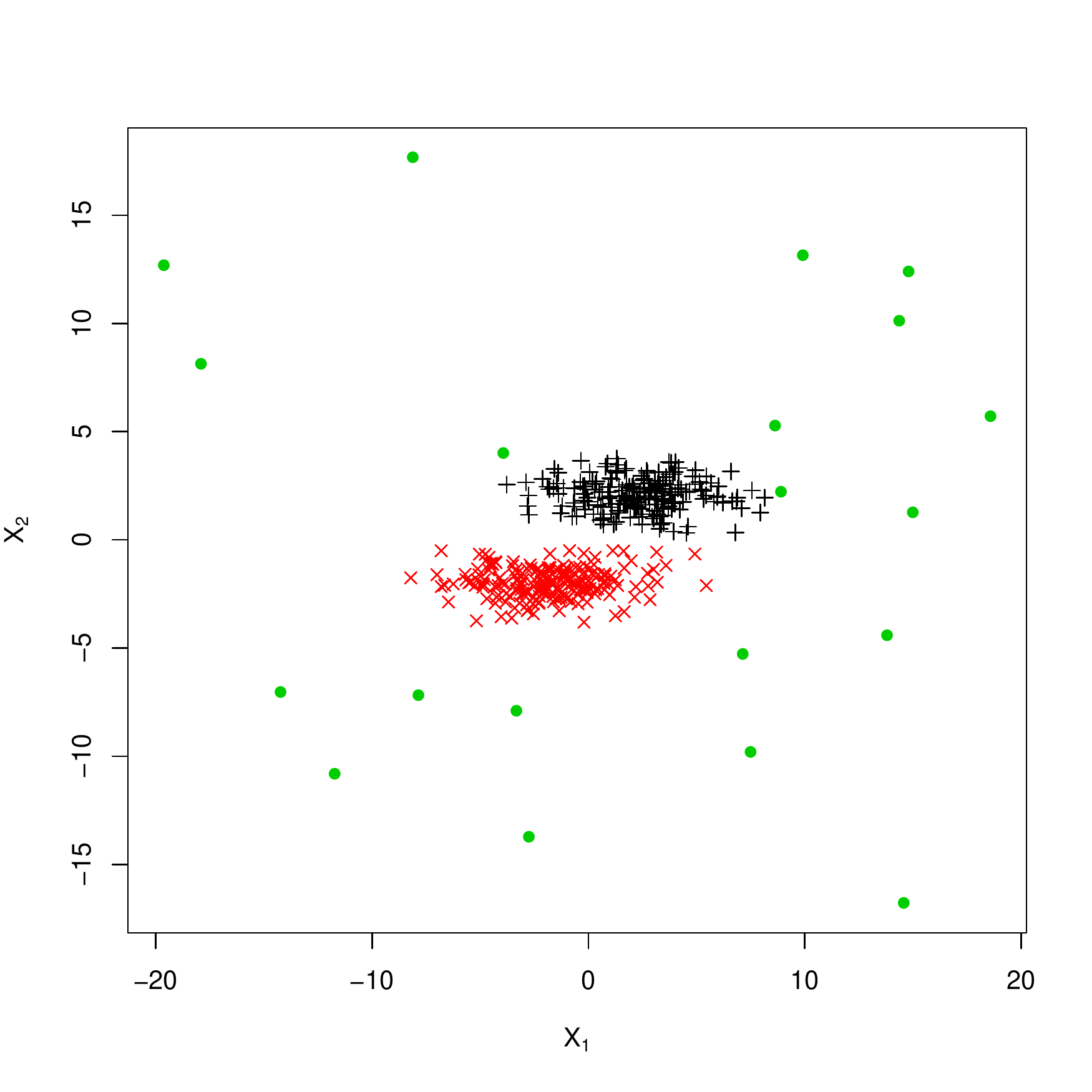}
\caption{
Scatterplot of the artificial data of Section~\ref{subsec:Artificial data}.
Noise is represented by green bullets.
\label{fig:art1}
}
\end{figure}

To illustrate the use of the package, we begin with an artificial data set from a mixture of $G=2$ bivariate normal distributions, of equal size, with an EEI structure for the component covariance matrices.
Twenty noise points are also added from a uniform distribution over the range -20 to 20 on each variable.
The data are generated by the following commands
\begin{CodeInput}
R> library("ContaminatedMixt")
R> library("mnormt")
R> p <- 2
R> set.seed(12345)
R> X1 <- rmnorm(n = 200, mean = rep(2, p), varcov = diag(c(5, 0.5)))
R> X2 <- rmnorm(n = 200, mean = rep(-2, p), varcov = diag(c(5, 0.5)))
R> noise <- matrix(runif(n = 40, min = -20, max = 20), nrow = 20, ncol = 2)
R> X <- rbind(X1, X2, noise)
\end{CodeInput}
The scatterplot of these data, in \figurename~\ref{fig:art1}, is obtained via the commands
\begin{CodeInput}
R> group <- rep(c(1, 2, 3), times = c(200, 200, 20))
R> plot(X, col = group, pch = c(3, 4, 16)[group], asp = 1, 
 +      xlab = expression(X[1]), ylab = expression(X[2]))
\end{CodeInput}

\subsubsection{Model-based clustering}
\label{subsubsec:Model-based clustering}
 
\begin{figure}[!ht]
\centering
\includegraphics{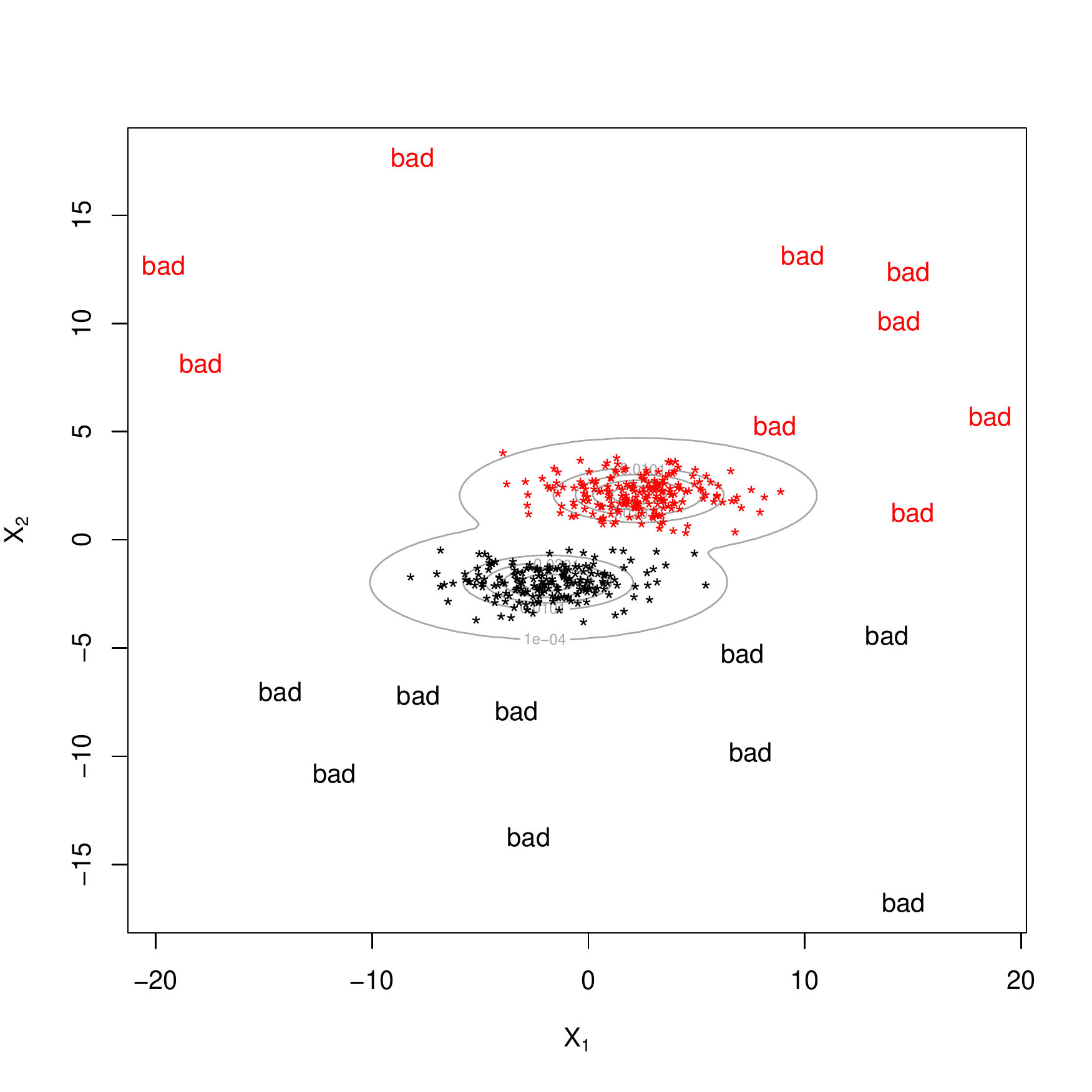}
\caption{
Clustering results from the model selected by the BIC on the artificial data of Section~\ref{subsec:Artificial data}.
Isodensities of the model are superimposed on the plot.
\label{fig:art2}
}
\end{figure}

We start with a model-based clustering analysis by considering all the fourteen models in \tablename~\ref{tab:models} and a number $G$ of clusters ranging from 1 to 4, resulting in 56 different models. 
The following command
\begin{CodeChunk}
\begin{CodeInput}
R> res1 <- CNmixt(X, model = NULL, G = 1:4, parallel = TRUE)  
\end{CodeInput}
\begin{CodeOutput}
With G = 1, some models are equivalent, so only one model from each set of 
equivalent models will be run.

 Using 8 cores


Best model according to AIC has G = 3 group(s) and parsimonious structure VVI 

Best model according to AICc, AICu, AIC3, AWE, BIC, CAIC, ICL has G = 2 
group(s) and parsimonious structure EEI
\end{CodeOutput}
\end{CodeChunk}
performs the ECM-fitting of the models and returns an object of class \code{ContaminatedMixt}. 
Because several models have to be fitted, parallel computation is convenient; it is set with the argument \code{parallel = TRUE}.
The number of CPU cores used is printed at video and it is followed, after a few seconds, by a description of the best model according to each of the 8 criteria in \tablename~\ref{tab:IC}.
Here, we can note as all the criteria, apart from the AIC, agree in suggesting a model with $G=2$ clusters and the true but unknown parsimonious structure EEI.
To find out more about the model selected by the BIC, we run the command
\begin{CodeChunk}
\begin{CodeInput}
R> summary(res1)
\end{CodeInput}
\begin{CodeOutput}
----------------------------------
Best fitted model according to BIC 
----------------------------------
 log.likelihood   n par   BIC
        -1835.8 420  11 -3738

Clustering table:
  1   2 
209 211 

Prior: group 1 = 0.4965, group 2 = 0.5035
Model: EEI (diagonal, equal volume and shape) with 2 components

Variables
 Means:
    group 1 group 2
X.1 -1.8564  2.3207
X.2 -1.9783  2.0697

 Variance-covariance matrices:
  Component 1
       X.1     X.2
X.1 5.0324 0.00000
X.2 0.0000 0.51525
  Component 2
       X.1     X.2
X.1 5.0324 0.00000
X.2 0.0000 0.51525

 Alpha
0.9506963 0.9485337

 Eta
86.44625 99.15542
\end{CodeOutput}
\end{CodeChunk}
As we can note from the estimates $\widehat{\eta}_1=86.44625$ and $\widehat{\eta}_2=99.15542$, there is a large enough degree of contamination in the two clusters, which together contributes to capture the underlying noise (see also the estimates $\widehat{\alpha}_1=0.9506963$ and $\widehat{\alpha}_2=0.9485337$).
In order to evaluate the agreement of the obtained clustering with respect to the true one, we can adopt the \code{agree()} function, included in the package, in the following way
\begin{CodeChunk}
\begin{CodeInput}
R> agree(res1, givgroup = group)
\end{CodeInput}
\begin{CodeOutput}
        groups
givgroup   1   2 bad points
       1   0 200          0
       2 200   0          0
       3   0   2         18
\end{CodeOutput}
\end{CodeChunk}
Apart from two bad points which are erroneously attributed to group 1, the obtained classification is in agreement with the true one.
Note that these misclassifications are not necessarily an error: the way the noisy points are inserted into the data makes possible that some of them may lay in the same range of good points and, as such, these points are detected as good points by the model.  
A plot of the clustering results for the best BIC model is displayed with the command (\figurename~\ref{fig:art2})
\begin{CodeInput}
R> plot(res1, contours = TRUE, asp = 1, xlab = expression(X[1]), 
 +      ylab = expression(X[2]))
\end{CodeInput}
Isodensities are also displayed (\code{contours = TRUE}).
Results in \figurename~\ref{fig:art2} look similar to those in \figurename~\ref{fig:art1}.

\subsubsection{Model-based classification}
\label{subsubsec:Model-based classification}

On the same data, we can also suppose to know the cluster membership of some of the available observations and evaluate the classification of the remaining ones. 
Via the commands
\begin{CodeChunk}
\begin{CodeInput}
R> indlab <- sample(1:400, 20)
R> lab <- group[indlab]
R> res2 <- CNmixt(X, G = 2, model = "EEI", ind.label = indlab, label = lab)
\end{CodeInput}
\begin{CodeOutput}
Estimating model EEI with G = 2:*********************************************
**********************************************


Estimated one model with G = 2 group(s) and parsimonious structure EEI
\end{CodeOutput}
\end{CodeChunk}
we firstly randomly select twenty good observations to be considered as labeled, and then we fit the EEI model (\code{model = "EEI"}), with $G=2$ clusters, assuming the groups membership of these observations as known in advance.
The position of the labeled observations is contained in the object \code{indlab}, while their group membership is given in the object \code{lab}. 
The agreement between the obtained classification and the true classification of the unlabelled observations only can be evaluated via the command
\begin{CodeChunk}
\begin{CodeInput}
R> agree(res2, givgroup = group)
\end{CodeInput}
\begin{CodeOutput}
   groups
      1   2 bad points
  1   0 193          1
  2 186   0          0
  3   0   0         20
\end{CodeOutput}
\end{CodeChunk}
Naturally, the comparison is automatically focused on the unlabelled observations.
As we can see, the results slightly change with respect to the clustering analysis and only one good observation from the first cluster is erroneously detected as bad.

\subsection[The wine dataset]{The \code{wine} dataset}
\label{subsec:The wine dataset}

\begin{figure}[!ht]
\centering
\resizebox{1\textwidth}{!}{
\includegraphics{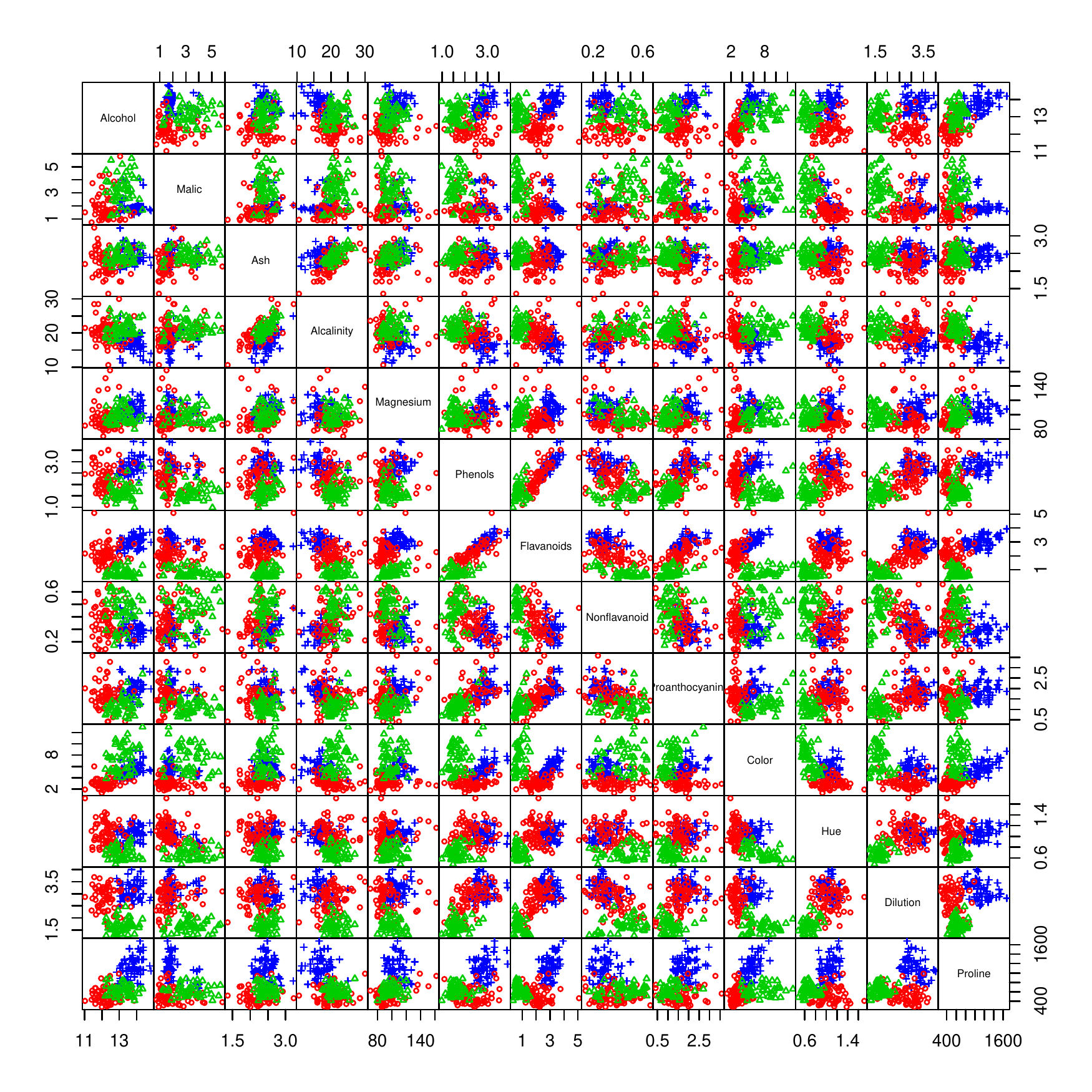}
}
\caption{
Wine data: Scatterplot matrix with clustering induced by the three cultivars.
\label{fig:wine1}
}
\end{figure}

This second tutorial uses the \code{wine} data set included in the \pkg{ContaminatedMixt} package and available at the UCI machine learning repository \url{http://archive.ics.uci.edu/ml/datasets/Wine}.
These data are the results of a chemical analysis of wines grown in the same region in Italy but derived from three different cultivars (Barbera, Barolo, and Grignolino). 
The analysis determined the quantities of $p=13$ constituents (continuous variables) found in each of the three types of wine. 
Data are loaded with
\begin{CodeInput}
R> data("wine")
\end{CodeInput}
This command loads a data frame with the first column being a factor indicating the type of wine and the others containing the measurements about the 13 constituents.
The plot of these data, displayed in \figurename~\ref{fig:wine1}, is obtained by
\begin{CodeInput}
R> group <- wine[, 1]
R> pairs(wine[, -1], cex = 0.6, pch = c(2, 3, 1)[group], 
 +       col = c(3, 4, 2)[group], gap = 0, cex.labels = 0.6)
\end{CodeInput}

\begin{figure}[!ht]
\centering
\resizebox{1\textwidth}{!}{
\includegraphics{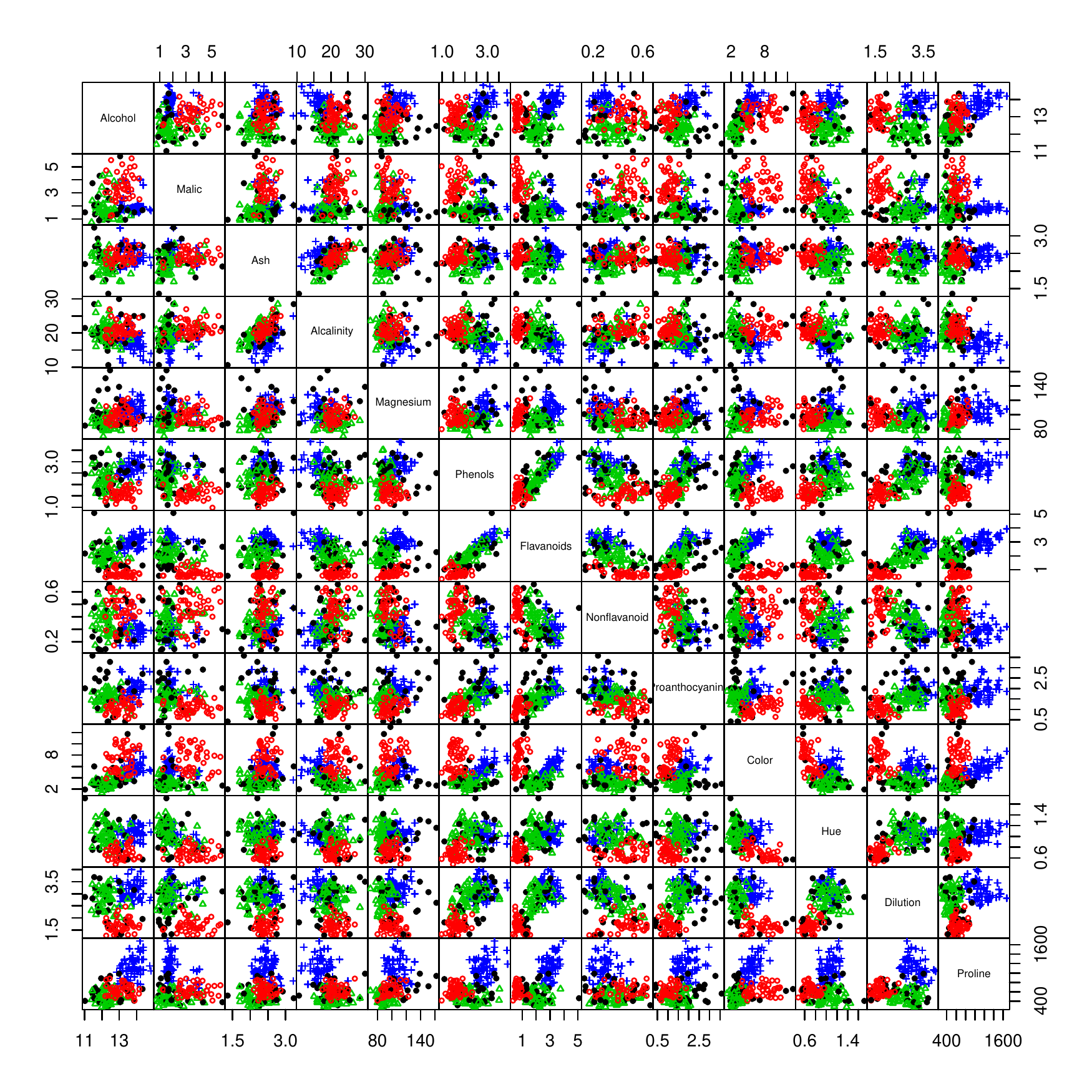}
}
\caption{
Wine data: Classification results from the model selected by BIC and ICL.
Bad points are represented by black bullets.
\label{fig:wine2}
}
\end{figure}

The command 
\begin{CodeChunk}
\begin{CodeInput}
R> res3 <- CNmixt(wine[, -1], G = 1:4, initialization = "random.soft", 
 +                seed = 5, parallel = TRUE)
\end{CodeInput}
\begin{CodeOutput}
With G = 1, some models are equivalent, so only one model from each set of 
equivalent models will be run.

 Using 8 cores


Best model according to AIC has G = 3 group(s) and parsimonious structure EVV 

Best model according to AICc has G = 4 group(s) and parsimonious structure 
EVE 

Best model according to AICu, CAIC has G = 3 group(s) and parsimonious 
structure EVI 

Best model according to AIC3 has G = 2 group(s) and parsimonious structure 
EVV 

Best model according to AWE has G = 3 group(s) and parsimonious structure EEI 

Best model according to BIC, ICL has G = 3 group(s) and parsimonious 
structure EEE 
\end{CodeOutput}
\end{CodeChunk}
fits all the fourteen models for $G\in\left\{1,2,3,4\right\}$, for a total of 56 models.
In this case, a random soft initialization is used (\code{initialization = "random.soft"}) with a pre-specified seed of random generation (\code{seed = 5}).
The best model, for the most commonly used criteria BIC and ICL, is EEE.
The classification performance of this model can be seen via the command
\begin{CodeChunk}
\begin{CodeInput}
R> agree(res3, givgroup = group)
\end{CodeInput}
\begin{CodeOutput}
             groups
givgroup      1  2  3 bad points
  Barbera    44  0  0          4
  Barolo      0  0 59          0
  Grignolino  0 49  0         22
\end{CodeOutput}
\end{CodeChunk}
As we can note, there are not misclassified wines; however, 26 wines are recognized as bad, 22 of which arise from the Grignolino cultivar.
Normal and $t$ mixtures do not allow this detection; moreover their classification, as estimated via the packages \pkg{mixture} and \pkg{teigen}, is not as good as that provided above \citep[cf.][]{Punz:McNi:Robu:2015}.
The graphical representation of the classification from the selected model can be obtained via the command (see \figurename~\ref{fig:wine2})
\begin{CodeInput}
R> pairs(res3, cex = 0.6, gap = 0, cex.labels = 0.6)
\end{CodeInput} 

\section{Conclusions}
\label{sec:conclusions}

In this paper, we have introduced \pkg{ContaminatedMixt}, a package for the \proglang{R} software environment, specifically conceived for fitting and disseminating parsimonious mixtures of multivariate contaminated normal distributions.
Although these models have been originally proposed for clustering applications \citep{Punz:McNi:Robu:2015}, their use has been here extended to model-based classification, where information about the group membership of some of the observations is available. 
The package is also meant to be a user-friendly tool for an automatic detection of mild outliers (also referred to as bad points herein).
Advantageously, computation can take advantage of parallelization on multicore PCs and computer clusters, when a comparison among different models is needed.
This is handy when, as it is often the case in practical applications, the number of clusters and/or the covariance structure of the model is not \textit{a priori} known. 
We believe our package may be a practical tool supporting academics and practitioners who are involved in robust cluster/classification analysis applications.

%
%
%
 %


\end{document}